\documentclass[12pt,preprint]{article}%
\usepackage{amssymb}
\usepackage{amsmath}
\usepackage{graphics}
\usepackage{epsfig}
\usepackage{amsfonts}
\usepackage{graphicx}%
\renewcommand{\vec}[1]{{\bf #1}}
\newcommand{\eqb}{\begin{equation}}
\newcommand{\eqe}{\end{equation}}
\newcommand{\dmb}{\begin{displaymath}}
\newcommand{\dme}{\end{displaymath}}

\newcommand{\eab}{\begin{eqnarray}}
\newcommand{\eae}{\end{eqnarray}}

\newcommand{\e}{\mbox{e}}
\newcommand{\be}{\begin{equation}}
\newcommand{\ee}{\end{equation}}

\begin{document}
\begin{titlepage}
\begin{flushright}
\end{flushright}
\vspace{0.6cm}
\begin{center}
\Large{Charged lepton spectra from hot-spot evaporation}
\vspace{1.5cm}\\
\large{Julian Moosmann$^\dagger$ and Ralf Hofmann$^*$}
\end{center}
\vspace{1.5cm}
\begin{center}

{\em $\mbox{}^\dagger$ Laboratorium f\"ur Applikationen der Synchrotronstrahlung\\
Universit\"at Karlsruhe (TH)\\ 
Postfach 6980\\ 
76128 Karlsruhe, Germany}

\end{center}
\vspace{1.5cm}
\begin{center}
{\em $\mbox{}^*$ Institut f\"ur Theoretische Physik\\
Universit\"at Heidelberg\\
Philosophenweg 16\\
69120 Heidelberg, Germany}
\end{center}
\vspace{1.5cm}
\begin{abstract}
Spectra for the emission of charged leptons from evaporating hot-spots
of preconfining phase in SU(2) Yang-Mills thermodynamics are
computed. Specifically, we consider charged single and dileptons with
their spectra being functions of energy and invariant mass,
respectively. In the former case, our results relate to narrow and
correlated electron and positron peaks measured in supercritical
heavy-ion collisions performed at GSI in the 1980ies. In the latter
case, we point out how strongly the spectra depend on typical
kinematic cuts (CDF analysis of Tevatron Run II data). We also propose
a scenario on how muon events of anomalously high multiplicity and
large impact-parameter modulus arise in the Tevatron data.

\end{abstract}
\end{titlepage}

\section{Introduction}

Pure SU(2) Yang-Mills thermodynamics possesses a narrow preconfining
phase characterized by condensed monopole-antimonopole pairs. The
latter originate as isolated defects through (anti)caloron
dissociation in the deconfining phase at higher temperatures
\cite{Hofmann2005,Hofmann2007}. Shortly above the critical temperature
$T_{\tiny\mbox{H}}$ for the Hagedorn transition towards the confining
phase the preconfining thermodynamics of the spatially infinitely
extended gauge-theory system is dominated by this thermal ground
state: The only propagating massive, dual gauge mode decouples, that
is, acquires a very large Meissner mass. This domination leads to
negative pressure of order $-T^4_{\tiny\mbox{H}}$. Once temperature
falls below $T_{\tiny\mbox{H}}$ the infinitely extended
monopole-antimonopole condensate decays nonthermally into single and
selfintersecting center-vortex loops (CVLs) which are interpreted as
spin-1/2 fermions \cite{Hofmann2005, Zpinch, burst, lowTemp}. Only
one-fold selfintersecting CVLs are absolutely stable in an SU(2)
Yang-Mills theory, see \cite{lowTemp,MH2008-1,MH2008-2}. Thus it is
these fermions that are associated with charged leptons and their
antiparticles. As a consequence, there is one SU(2) Yang-Mills theory
for each lepton family with the respective Yang-Mills scale $\Lambda$
matching the mass of the charged lepton: $\Lambda_e\sim
m_e\,,\Lambda_\mu\sim m_\mu\,,\Lambda_\tau\sim m_\tau\,.$ This
scenario, although at first sight ruled out by the apparent
experimentally inferred pointlikeness of charged leptons, has
predictive power under conditions where the creation of large-sized
hot-spots of preconfining ground state is feasible. These conditions
include very high energies in particle collisions \cite{burst}, high
temperatures in extended systems \cite{Zpinch} and/or local energy
densities exceeding $m_e^4\sim 10^{11}\,$keV$^4$. Recall that,
according to SU(2) Yang-Mills thermodynamics, the apparent
pointlikeness of leptons when probed by high-momentum transfer
Standard-Model (SM) processes is a consequence of the presence of a
large number of instable excitations in the interaction region. Their
combined effect together with the free shiftability of vortex
intersection points is efficiently and successfully described
perturbatively by the SM. In this context, a succession of vertices
and off-shell propagators (Feynman diagrams) mediates the
transformation of structureless initial into the according final
states.

Recently, anomalous, high-multiplicity muon events of {\sl
symmetrically} (positive-negative) distributed and {\sl power-like}
decaying (as opposed to exponentially cut-off) rate as a function of
impact factor $R$ were reported to occur in high-energy particle
collisions \cite{CDF}. These data are generated by protons and
antiprotons colliding head-on at $\sqrt{s}=1.96\,$TeV (Tevatron Run
II). The invariant mass spectrum of dimuons created outside the vacuum
of the beam pipe exhibits anomalous structure in the GeV region.  Also,
taking an additional muon into account, the according experimental
rate is not explicable by any known SM processes. We are aware of the
fact that the experimental situation in the case of anomalous
multimuon events produced at the Tevatron as of yet is not clear cut
and contested \cite{CDF,D0note,DorigoBlog}. Motivated by existing experiments 
\cite{CDF,D0note}, our theoretical dimuon spectra, obtained by investigating SU(2)$_\mu$
hot-spot evaporation under certain kinematic cuts, are intended to offer guidelines for
future analysis of collider date in the TeV realm.

We also consider single-particle spectra for $e^+$, $e^-$ emission
from evaporating SU(2)$_e$ hot spots. Here the motivation arises due
to the observation of universal narrow-peak structures in
supercritical heavy-ion collisions at Gesellschaft f\"ur
Schwer\-ionen\-forschung Darmstadt (GSI) more than two decades ago.

\subsection{Experimental situation: Multilepton events\label{Intro}}

{\sl Electron and positron emission in supercritical heavy-ion
  collisions (GSI).}  In supercritical heavy-ion collision systems
  such as U+Th, Th+Th, Th+Cm, U+U, U+Cm of combined nuclear charge
  $Z_u$ between 180 and 188 and thus well above criticality at
  $Z_{u,c}=173$ the emission of narrow positron peaks of several ten
  keV width and center at 250\,keV, 330\,keV, and 400\,keV was
  detected \cite{Greiner,PRL1,PRL2}. These narrow peaks sit on a broad
  background which is explained by conventional, parameter-free
  theory. Later the correlated emission of narrow peaks of electrons
  with similar characteristics was observed \cite{PRL3}. Both
  signatures decisively were ruled out to be due to any nuclear
  reactions when bombarding energies near the Coulomb barrier are
  applied. The emergence of these narrow peaks was theoretically
  approached by postulating large delay times of the order $3\times
  10^{-21}\,$s which, however, are hard to reconcile with the smaller
  average delay time in deep inelastic reactions, see \cite{Greiner}
  and references therein.

The width of the narrow $e^+, e^-$ peaks in sum energy is narrower
than that of the single positron lines. Also, from an analysis of
Doppler broadening one concludes that the emitting source moves very
slowly ($v/c\le 0.05$). At the same time, compared to the situation of
a weak constraint on opening angle (between $40$ and $170$ degrees) no
sizable detector activity was seen in the search for correlated
back-to-back emission of electrons and positrons.  The combination of
the latter two signatures excludes the {\sl pair} (two-particle) decay
of a neutral particle state in the mass range between 1.5\,MeV to
2\,MeV which from an analysis of the invariant mass spectrum alone
would not be excluded!

Furthermore, the prediction of a strong dependence of peak energies
and intensities on the combined nuclear charge $Z_u$, derived from the
hypothesis of a long-lived, giant dinuclear complex needed for
spontaneous pair creation\footnote{Strong-field QED, local decay of
the QED vacuum: occupation of the $1s$ state of binding energy (use of
natural units $c=\hbar=k_B=1$) $E_b\sim-2m_e$ by a state of equal energy
transferred from the Dirac sea.}, does not match with the experimental
data. Rather, the data point to an independence of the peak features
on $Z_u$. This universality is a puzzle in conventional strong-field
QED which predicts strong power laws in $Z_u$.

Considering all signatures mentioned above (large delay times
$\rightarrow$ formation of spatial regions with negative pressure
pointing to the existence of droplets of preconfining ground state of
an SU(2) Yang-Mills theory of scale $T_{\tiny\mbox{H}}\sim
m_e=511\,$keV; no pair decay but still a heavy neutral primary
$\rightarrow$ droplets of energy content several times $m_e$
evaporating isotropically; universality of peak energies and widths in
single positron and pair spectra $\rightarrow$ again, new phase of
matter, preconfining-phase droplets whose decay is insensitive to the
physics of their creation) we tend to attribute them to the formation
and subsequent evaporation of SU(2)$_e$ preconfing-phase hot-spots
whose one-particle thermal emission spectrum is simply computed, see
Sec.\,\ref{OPS}.

Let us now estimate the typical energy density associated with the
Coulomb fields at closest approach of the two nuclei. Notice that the
spontaneous creation of positrons in strong-field QED considered in
explaining the anomalously high yield \cite{Reinhardt} requires
electronic binding energies $E_b\sim -2\,m_e$. Since $Z_u \alpha\sim
1$ the associated Bohr radius $a_0$ is $a_0\sim |E_b|^{-1}$ , and the
maximal energy density $\rho_p$ in supercritical heavy-ion collision
is $\rho_p\sim a_0^{-4}\sim 10\,m_e^4$ which is larger (but not
hierarchically larger!) than what is minimally needed to traverse the
Hagedorn transition in SU(2)$_e$. Thus preconfining-phase hot-spots
likely are generated in the form of {\sl small} droplets (spatial
extent comparable with $a_0$).

In connection with the smallness of the droplets, an important
question concerns the narrowness of the observed peaks. Notice that
the observed width of the electron and/or positron peaks corresponds
to a lifetime of about $10^{-20}\,$s. This, however, is the predicted
lifetime for SU(2)$_e$ preconfing-phase hot-spot evaporation (with a
slow dependence on the deposited energy), see Eq.\,(\ref{lifetimes})
below. If the energy deposited during the formation of a hot-spot is
larger but comparable to the peak energy of the single-particle
spectrum for {\sl thermal} electron or positron emission\footnote{By
thermal we mean that the hot-spot is considered sufficiently large to
{\sl not} wobble about indeterministically. That is, one can assume a
smoothly shrinking spherical surface due to recoil-free evaporation.}
then the hot-spot and its decay signatures must be interpreted quantum
mechanically. The droplet's thus {\sl computed} lifetime \cite{burst},
however, still should coincide with the inverse width seen in the {\sl
experimental} decay spectra. From the fact that the observed peaks are
narrow in contrast to what is seen in the computed spectra we would
again conclude that the experimental hot-spots are small droplets with
a position uncertainty characterized by typical recoils. That is,
their mass is only several times larger than $m_e$. This is compatible
with the argument on available energy density for hot-spot creation
given above.

For the case, where the deposited energy in hot-spot creation would
exceed $m_e$ by many orders of magnitude, the observed peaks should
broaden, and asymptotically the experimental one-particle spectra
should match those computed in Sec.\,\ref{OPS}.\vspace{0.3cm}\\ {\sl
Multimuon events at large impact parameter (Tevatron Run II).}  For
anomalous multimuon events ignited by proton-antiproton primary
collisions at $\sqrt{s}=1.96\,$TeV (Tevatron Run II, total integrated
luminosity 2100\,pb$^{-1}$) and analyzed by the CDF collaboration
(however, not verified by DO \cite{D0note} who seem to apply a much
tighter constraint on the range of investigated impact parameter
\cite{DorigoBlog}), we are interested in the large number of dimuon
events originating outside the beam pipe. Namely, an anomalous rate of
dimuons, nearly independent of the sign of the impact parameter $R$
was detected for $|R|>1.5\,\mathrm{cm}=|R_{\tiny\mbox{BP}}|$.  Notice
that known QCD processes account for the dimuons created within the
vacuum of the beam pipe while the counts for dimuons originating from
vertices with $|R|>|R_{\tiny\mbox{BP}}|$ are in excess to SM
predictions.

The motivation and precise implementation philosophy of
Ref.\,\cite{CDF} in imposing certain kinematic cuts during data
analysis remains largely obscure to the present authors. In part this
is explicable by a profound lack of experience in addressing the
complex data situation posed by TeV range primary
collisions. Nevertheless, we would like to point out what our
evaporating hot-spot model predicts about the shape of the dimuon
invariant mass spectrum at large impact-parameter modulus, see below,
when applying typical kinematic constraints chosen by the
experimentalists.

In \cite{CDF} a subset of the Run-II data, subject to certain
kinematic constraints in transverse momentum and opening angle
(integrated luminosity now only 742\,pb$^{-1}$) was analyzed, and
so-called ghost events were isolated. These events share the following
features: neither the approximate symmetry under $R\to -R$ and the
decay of the rate distribution of multimuon events with increasing
$|R|$, nor the large magnitude of the rate of events containing two or
more muons for $|R|>|R_{\tiny\mbox{BP}}|$, nor the distribution of
rate for a given charge composition of dimuon events in invariant mass
is explained by known SM processes. Probably depending on the imposed
kinematic cuts ($p_T\ge 2\,$GeV) the measured invariant mass spectra
of dimuons at large $|R|$ typically have large weight in the GeV
region. To link this to our SU(2)$_\mu$ approach, we will in
Sec.\,\ref{DMG} present a number of theoretical results on dimuon
spectra based on a single evaporating SU(2)$_\mu$ hot-spot when
varying kinematic cuts.

Since such a calculation unrealistically assumes an impact-parameter
independence of the process creating the hot-spot and since it ignores
the physics of prerequisite processes neither the impact-parameter
dependence nor the overall normalization of the dimuon spectra are
predicted here. For a prediction of these two features additional
deliberations and assumptions must be made. It is clear, however, that
a preconfining SU(2)$_\mu$ hot-spot of total mass $\sim 1\,$TeV is
almost at rest and thus emits muons {\sl isotropically} explaining the
observed approximate symmetry under $R\to -R$ in the impact-parameter
distribution of the experimental yields. (Emission towards the beam
pipe is as likely as away from it at given distance.)

We tend to believe that the reason why QCD predictions and
experimental data agree well for events originating at
$|R|<|R_{\tiny\mbox{BP}}|$ is the absence of heavy nuclei whose strong
electric fields trigger hot-spot creation assisted by the rare TeV
photons of electromagnetic $p\bar{p}$ annihilation.  Outside the beam
pipe the detector material provides for these nuclei. Assuming the
probability for the conversion of a TeV photon into an SU(2)$_\mu$
hot-spot inside the Coulomb potential of a heavy nucleus to be
sufficiently small\footnote{In principle, this is calculable under
specific assumptions. For example, one could presume that inside the
Coulomb potential of the nucleus the incoming photon perturbatively
creates a single pair of TeV dimuons which in turn originate a
perturbative cascade producing an ever increasing number of low-energy
muons until their number density reaches criticality for the Hagedorn
transition.}, the almost unadulterated total photon flux decreases
(geometrically) in a power-like way as a function of $R$ while the
evaporation physics is independent of $R$. This would explain the
slow, power-like decay of multimuon activity at large $|R|$.

This article is organized as follows: In the next section we compute
the spectrum for emission of single charged leptons, arising from
evaporation of static SU(2) hot-spots, as a function of lepton
energy. Specifically for SU(2)$_e$, we also determine hot-spot life
times in dependence of the deposited energy and observe agreement with
the inverse width of the narrow peaks detected at GSI. In
Sec.\,\ref{DMG} we compute the spectra for the emission of charged
dileptons from the evaporation of static SU(2) hot-spots as a function
of the pair's invariant mass and under various kinematic cuts. As a
result, these spectra sensitively depend on the kinematic cuts
applied. Sec.\,\ref{S&C} discusses our results in view of present and
future analysis of TeV-range collider data.

\section{Single lepton spectrum from hot-spot evaporation\label{OPS} }

For temperatures just above $T_{\tiny\mbox{H}}$ the SU(2) Yang-Mills
system is ground-state dominated. Once the creation of a bubble of
radius $r_0$ containing this new phase of matter has taken place its
evaporation is simply determined by the content of stable, final
particle species (number of selfintersections in center-vortex loop,
spin, and charge), energy conservation, and the value of the Hagedorn
temperature $T_{\tiny\mbox{H}}$.

Since we are interested in leptons $l$ far away from the emitting
hot-spot surface we can address the problem of calculating their
spectra thermodynamically setting
$T=T_{\tiny\mbox{H}}\equiv\frac{11.24}{2\pi}y\,m_{l=1}$ ($l=1$ is the
charged lepton state of the SM) where $y$ is of order unity expressing
a theoretical uncertainty. In the following we strongly appeal to the
results of \cite{burst}.

Leaving the angular integration in Eq.\,(7) of \cite{burst}
explicit\footnote{Microscopically, Lambert's law states that at a
given point of the emitting surface only the momentum component
perpendicular to this surface gets Fermi-distributed at
$T_{\tiny\mbox{H}}$ far away from the hot-spot system and thus is
measurable. Thus the factor 1/4 in the first half of Eq.\,(7) of
\cite{burst} actually should be replaced by unity.} and setting $l=1$,
the number $\eta_{l=1}$ of charged leptons emitted per unit time and
surface reads
\begin{equation}
  \label{intnumberflux}
  \eta_{l=1}=\int_{0}^{\infty}dp\int_{-1}^1 d\cos\theta \int_0^{2\pi}d\varphi\,\frac{M_{l=1}}{8\pi^{3}\sqrt{p^{2}
      +m_{l=1}^{2}}}\frac{p^{3}}{e^{\sqrt{p^{2}+m_{l=1}^{2}}/T_{\tiny\mbox{H}}}+1}\,,
\end{equation}
where $M_{l=1}=2\times 2=4$ (2-fold spin degeneracy, two-fold charge
degeneracy\footnote{Because of the large reservoir of to-be-emitted
charged leptons in a hot-spot and the assumed absence of external
fields the evaporation physics is charge and spin-orientation
blind. In reality, this can not entirely be true because of a possible
fluctuating, CP violating axion field (whose homogeneous incarnation
is of cosmological relevance \cite{GiacosaHofmann2005}) that slightly
prefers the emission of negative over positive charge. However, for
the total yields this effect is negligible.}, $p=|\vec{p}|$ is the
modulus of the spatial on-shell momentum of a single charged lepton
far away from the emitting surface, and $m_{l=1}$ can be set equal to
$m_e=511\,$keV for SU(2)$_e$ and $m_{\mu}=105.6\,$MeV for
SU(2)$_\mu$. From Eq.\,(\ref{intnumberflux}) we have for the
differential yield $\frac{d \eta_{l=1}}{dp}$ per unit on-shell
momentum, time and surface
\eqb\label{diffyield} \frac{d \eta_{l=1}}{dp}=\frac{2}{\pi^{2}\sqrt{p^{2}
    +m_{l=1}^{2}}}\frac{p^{3}}{e^{\sqrt{p^{2}+m_{l=1}^{2}}/T_{\tiny\mbox{H}}}+1}\,,
\eqe
or alternatively, the differential yield $\frac{d\eta_{l=1}}{dE}$ per unit
on-shell energy $E$, time and surface
\eqb\label{diffyieldE} \frac{d
  \eta_{l=1}}{dE}=\frac{2}{\pi^{2}}\frac{(E^2-m_{l=1}^2)}{e^{E/T_{\tiny\mbox{H}}}+1}\,.
\eqe
To obtain the differential number $dN_{l=1}/dE$ of emitted charged leptons per
bin of energy we multiply $\frac{d\eta_{l=1}}{dE}$ by the actual surface $4\pi
r(t)^2$ of the spherical hot-spot and subsequently integrate over time $t$
from zero to $t_{ev}$ \cite{burst}:
\eqb
\label{diffnumb}
\frac{dN_{l=1}}{dE}=\left(4\pi \int_0^{t_{ev}} dt\, r(t)^2\right)\times\frac{d
  \eta_{l=1}}{dE}\,.  \eqe
According to Eq.\,(5) of \cite{burst} one has
\begin{equation}
  r(t)=r_{0}-\frac{J(T_{{\tiny \mbox{H}}})}{\rho_{{\tiny \mbox{H}}}}\,t\,,
\end{equation} 
and according to Eq.\,(6) of \cite{burst}
\begin{equation}
  \label{Liftumus}
  t_{ev}=\frac{\rho_{{\tiny \mbox{H}}}}{J(T_{{\tiny\mbox{H}}})}r_{0}\,,
\end{equation}
where
\eqb
\label{defs}
\rho_{\tiny\mbox{H}}=22.48\,y^{4}m_{l=1}^{4}\,,\ \ \
r_{0}=\left(\frac{3}{4\pi}\frac{\delta\sqrt{s}}{\rho_{{\tiny \mbox{H}}}
  }\right)^{1/3}\,,\ \ \
J(T_{\tiny\mbox{H}})=J_{l=0}(T_{\tiny\mbox{H}})+J_{l=1}(T_{\tiny\mbox{H}})\,,
\eqe
and
\begin{equation}
  \label{Jexp}
  J_{l}(T_{\tiny\mbox{H}})=M_l \int_{0}^{\infty}dp\,
  \frac{1}{2\pi^{2}}\frac{p^{3}}{e^{\sqrt{p^{2}+m_{l}^{2}}/T_{\tiny\mbox{H}}}+1}\,.
\end{equation}
In Eq.\,(\ref{Jexp}) one has $m_0=0$ and $M_0=2$ (neutrinos are Majorana), in
Eq.\,(\ref{defs}) $\delta\sqrt{s}$ represents the fraction of c.m. energy
going into hot-spot creation ($\delta$ of order unity), and
$y\equiv\frac{\Lambda}{m_{l=1}}$ where $\Lambda$ is the Yang-Mills scale of
the SU(2) gauge theory. This implies
$T_{\tiny\mbox{H}}=\frac{11.24}{2\pi}y\,m_{l=1}$, see for example
\cite{GiacosaHofmann2005}.

\begin{figure} [ptb]
  \begin{center}
    \includegraphics[width=0.8\textwidth ] {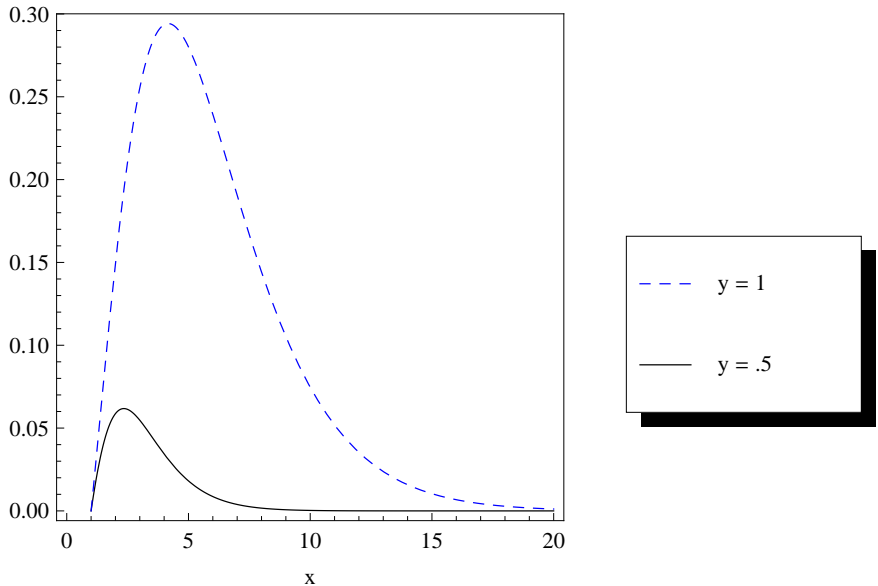}
    \caption{Dimensionless one-particle spectral shapes $S_{l=1}$ for
      the emission of charged leptons of dimensionless energy
      $x=E/m_{l=1}$ due to SU(2) hot-spot evaporation for $y=1/2$
      (solid) and $y=1$ (dashed).}
    \label{Fig-1}
  \end{center}
\end{figure}

The typical inverse width
\eqb
\label{narrowwidth}
\Gamma^{-1}\sim \frac{1}{70\,\mbox{keV}}=9.4\times 10^{-21}\,\mbox{s} \eqe
of single electron or positron peaks seen in supercritical heavy-ion
collisions \cite{Greiner,PRL1,PRL2,PRL3} compares well with the life time
$t_{ev}$ of hot-spots belonging to SU(2)$_e$ calculated by assuming
recoil-free, that is, thermal emission. Recall that by Eqs.\,(\ref{Liftumus}) and (\ref{defs})
$t_{ev}$ varies rather weakly with $\sqrt{s}$. Setting $\delta=y=1$, we have
\eab
\label{lifetimes}
\sqrt{s}&=&10\,\mbox{MeV}\Rightarrow t_{ev}=1.0\times
10^{-21}\,\mbox{s};\ \ \sqrt{s}=100\,\mbox{MeV}\Rightarrow
t_{ev}=2.15\times 10^{-21}\,\mbox{s};\nonumber\\
\sqrt{s}&=&1\,\mbox{GeV}\Rightarrow t_{ev}=4.63\times
10^{-21}\,\mbox{s};\ \ \ \ \sqrt{s}=10\,\mbox{GeV}\Rightarrow
t_{ev}=1.0\times 10^{-20}\,\mbox{s};\nonumber\\
\sqrt{s}&=&100\,\mbox{GeV}\Rightarrow t_{ev}=2.15\times
10^{-20}\,\mbox{s};\ \ \sqrt{s}=1\,\mbox{TeV}\Rightarrow
t_{ev}=4.63\times 10^{-20}\,\mbox{s}\,.\nonumber\\ \eae
Notice that according to Eq.\,(\ref{diffnumb}) the quantities $\delta$ and
$\sqrt{s}$ enter only into the normalization of the spectrum and not into its
shape.  For reasons of better interpretability it is advantageous to factor
out the dimensionful quantity $c_{l=1;y,\delta} m_{l=1}^{-1}$ where
$c_{l=1;y,\delta}$ denotes the number obtained by rescaling $E$ and
$T_{\tiny\mbox{H}}$ by $m_{l=1}^{-1}$ in the first factor of Eq.\,(\ref{diffnumb}). Explicitely,
we have
\begin{equation}
  \label{cs}
  c_{l=1;y,\delta}=
  \pi^2\delta\sqrt{\frac{s}{m_{l=1}^2}}\,
  \left[  \int_0^\infty d\tilde{x}  \,\tilde{x}^3
    \left(\frac{1}{\e^{\gamma\tilde{x}}+1}
      +\frac{2}{\e^{\gamma\sqrt{\tilde{x}^2+1}}+1}\right)\right]^{-1}\,,
\end{equation}
where $\gamma\equiv \frac{2\pi}{11.24\,y}$. For SU(2)$_e$ and
$\sqrt{s}=10\,$MeV this yields
\begin{align}
\label{csexp}
c_{l=1;y=1,\delta=1/2}&=0.569661\,,\ \ \ \ \ c_{l=1;y=1,\delta=1}=1.13932\,,\nonumber\\
c_{l=1;y=1/2,\delta=1/2}&=9.88092\,,\ \ \ \ \ c_{l=1;y=1/2,\delta=1}=19.7618\,.
\end{align}
The remaining factor in Eq.\,(\ref{diffnumb}) then is a dimensionless function $S_{l=1}$ of its
dimensionless arguments $x=E/m_{l=1}$ and $y$:
\eqb
\label{diffnumbfac}
\frac{dN_{l=1}}{dE}=c_{l=1;y,\delta}\,\,m_{l=1}^{-1}\times S_{l=1}(x,y)\,.
\eqe
In Fig.\,{\ref{Fig-1}} the function $S_{l=1}(x,y)$ is plotted in dependence of
$x$ for $y=1/2$ and $y=1$.
Notice the broad spectral shape of $S_{l=1}$. As mentioned in
Sec.\,\ref{Intro}, this is a consequence of our assumption that the
hot-spot can be treated as a classical, recoil-free object during the
entire history of its thermal evaporation. Obviously, this assumption
breaks down when the mass of the hot-spot becomes comparable to
$m_{l=1}$: The emission of charged leptons then is subject to a
quantum mechanical decay with the width of the latter still being
described by the life time computed from recoil-free, thermal
emission.

For the example of SU(2)$_e$, at $\sqrt{s}=10\,$MeV and at $\delta=1$,
Eq.\,(\ref{csexp}) and Fig.\,\ref{Fig-1} tell us that the number
$N^{\tiny\mbox{max}}_{l=1}$ of single positrons {\sl and} electrons
per energy bin of $m_e=511\,$keV and emitted at the spectral maximum
is
\eab
\label{maxnumb}
N^{\tiny\mbox{max}}_{l=1}(\delta=1,y=1)&=&1.14\times 0.3=0.342\,,\nonumber\\
N^{\tiny\mbox{max}}_{l=1}(\delta=1,y=1/2)&=&19.8\times 0.064=1.27\,.  \eae
Appealing to the number of counts per energy bin in the experimental
data \cite{PRL1,PRL2,PRL3}, the numbers in Eq.\,(\ref{maxnumb}) can be
used to obtain an estimate for the total number of hot spots created
in supercritical heavy-ion collisions at given integrated luminosity.

\section{Dilepton spectrum from hot-spot evaporation\label{DMG} }

Let us now derive predictions for the shape of the invariant mass spectrum of
dileptons emitted from the surface of an evaporating hot-spot.  Under certain
kinematic constraints this is the situation relevant to the analysis presented
in \cite{CDF}.  Considering a charged lepton far away from the hot-spot
surface and thus in thermal equilibrium, it is reasonable to assume that it is
not temporally and spatially correlated to another charged lepton emerging
from the same hot-spot. Moreover, it is experimentally impossible to resolve
the hot-spot spatially or temporally at presently available
energies\footnote{For electrons/positrons and $\sqrt{s}\sim 1\,$TeV the
  evaporation time is $\sim 10^{-19}$\,s and the initial radius $10^{-10}$\,m
  \cite{burst}. For muons/antimuons and comparable values of $\sqrt{s}$ these
  time and distance scales are even smaller. Since the CDF II single-hit
  resolution is $10^{-6}\,$m an evaporating hot-spot acts like a pointlike
  vertex experimentally.}. Thus, the entire hot-spot evaporation is seen as an
instant of extremely high-multiplicity emission of charged leptons emerging
from a point.

Having said the above, it is clear that the number of charged lepton pairs
$dN_{2(l=1)}$ with their members in momentum bins $dp_1$ and $dp_2$ and
angular bins $d\varphi_1$, $d\varphi_2$, $d\cos\theta_1$, and $d\cos\theta_2$
reads:
\eab
\label{dN}
dN_{2(l=1)}&=&\left(4\pi \int_0^{t_{ev}} dt\, r(t)^2\right)^2\times dp_1\,
dp_2\,
d\cos\theta_1\, d\varphi_1\, d\cos\theta_2\, d\varphi_2\times \nonumber\\
& &\frac{1}{4\pi^{6}\sqrt{p_1^{2}
    +m_{l=1}^{2}}}\frac{p_1^{3}}{e^{\sqrt{p_1^{2}+m_{l=1}^{2}}/T_{\tiny\mbox{H}}}+1}
\frac{1}{\sqrt{p_2^{2}
    +m_{l=1}^{2}}}\frac{p_2^{3}}{e^{\sqrt{p_2^{2}+m_{l=1}^{2}}/T_{\tiny\mbox{H}}}+1}\,.\nonumber\\ 
\eae
For the invariant mass $M$ of a charged lepton pair we have
\eqb
\label{invM}
M^2=2(m_{l=1}^2+\sqrt{m_{l=1}^2+p_1^2}\sqrt{m_{l=1}^2+p_2^2}-p_1
p_2\cos\theta)\,, \eqe
where $\theta\equiv\angle(\vec{p_1},\vec{p_2})$.  Solving Eq.\,(\ref{invM})
for $p_1$, the only acceptable solution is
\eab
\label{p1}
&&p_1(M,p_2,\cos\theta)=\nonumber\\
&&\frac{(M^2-2m_{l=1}^2)p_2\cos\theta+\sqrt{(m_{l=1}^2+p_2^2)
    (M^4-4M^2m_{l=1}^2-4(1-\cos^2\theta)m_{l=1}^2p_2^2)}}{2(m_{l=1}^2+(1-\cos^2\theta)p_2^2)}\,.\nonumber\\ 
\eae
Moreover, we have
\eqb
\label{costh}
\cos\theta=\cos\theta
(\theta_1,\theta_2,\varphi_1,\varphi_2)=\frac{\vec{p}_1\cdot\vec{p}_2}{p_1
  p_2}\,.  \eqe
Appealing to Eqs.\,(\ref{dN}), (\ref{p1}), and (\ref{costh}), we have for the
number $dN_{2(l=1)}$ of lepton pairs per invariant mass bin $dM$
\eab
\label{N}
\frac{dN_{2(l=1)}}{dM}&=&\left(4\pi \int_0^{t_{ev}} dt\, r(t)^2\right)^2\times
\int dp_2\,\frac{dp_1}{dM}\,d\cos\theta_1\, d\varphi_1\, d\cos\theta_2\, d\varphi_2\times \nonumber\\
&
&\left.\frac{1}{4\pi^{6}\sqrt{p_1^2+m_{l=1}^{2}}}\frac{p_1^{3}}{e^{\sqrt{p_1^{2}+m_{l=1}^{2}}/T_{\tiny\mbox{H}}}+1}
  \frac{1}{\sqrt{p_2^{2}
      +m_{l=1}^{2}}}\frac{p_2^{3}}{e^{\sqrt{p_2^{2}+m_{l=1}^{2}}/T_{\tiny\mbox{H}}}+1}\right|_{\tiny\begin{array}{c}
    p_1(M,p_2,\cos\theta);\\
    \cos\theta (\theta_1,\theta_2,\varphi_1,\varphi_2)\end{array}}\,,\nonumber\\
\eae
where the integration over $p_2$, $\theta_i$, and $\varphi_i$ is subject to
kinematic constraints. For example, towards the infrared one may cut off the
modulus of each lepton's momentum component $p_{i,\perp}$ ($i=1,2$) in the
plane tranverse to the beam direction. Moreover, the opening angle $\theta$ of
the lepton pair may be constrained by $1\ge\cos\theta\ge\cos\theta_0$ where
$\cos\theta_0$ is a fixed number between 1 and $-1$. According to
Eq.\,(\ref{costh}) the kinematic constraint on $\cos\theta$ implies
constraints on $\theta_i$ and on $\varphi_i$.

As in Sec.\,\ref{OPS} it is advantageous to rescale $p_2$,
$M$, and $T_{\tiny\mbox{H}}$ by $m_{l=1}^{-1}$ in Eq.\,(\ref{N}) and to factor the result in
analogy to Eq.\,(\ref{diffnumbfac}) into the following form
\eqb
\label{diffnumbfacl2}
\frac{dN_{2(l=1)}}{dM}=c_{2(l=1);y,\delta}\,\,m_{l=1}^{-1}\times
S_{2(l=1)}(X,y)\,,  \eqe
where $c_{2(l=1);y,\delta}\equiv c^2_{l=1;y,\delta}$ and
$X\equiv\frac{M}{m_{l=1}}$. Constraints on the transverse momentum are
then expressed in terms of the dimensionless version
$\pi_{i,\perp}\equiv\frac{p_{i,\perp}}{m_{l=1}}$.  The integral
associated with $S_{2(l=1)}(X,y)$ is performed using Monte Carlo
methods.

\begin{figure} [ptb]
  \begin{center}
    \includegraphics[width=1\textwidth,height=.42\textheight] {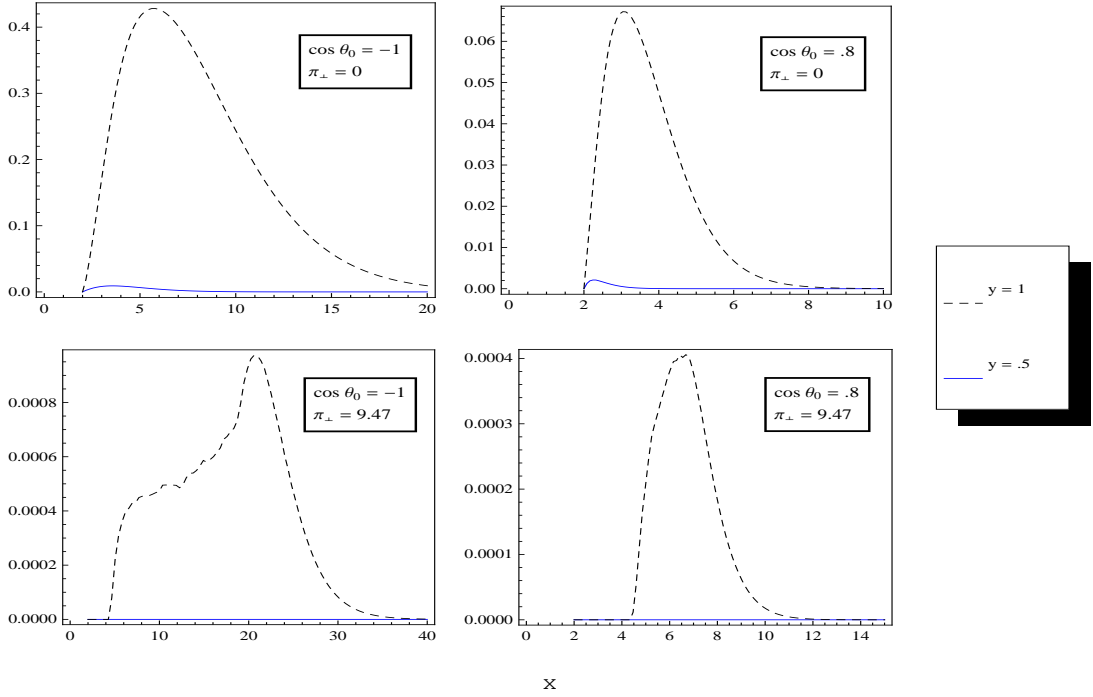}
    \caption{Plots of the function $S_{2(l=1)}(X,y)$ (dimensionless spectral
      distribution in invariant mass of charged dilepton events) at
      $\cos\theta_0=-1;0.8$, $\pi_{1,\perp}=\pi_{2,\perp}=0;9.47$, and
      $y=1/2;1$ as a function of $X$.}
    \label{Fig-2}
  \end{center}
\end{figure}

In Fig.\,\ref{Fig-2} plots of $S_{2(l=1)}(X,y)$ are presented as a function of
$X$ at $y=1/2$ and $y=1$, for $\cos\theta_0=-1;0.8$ (latter cut chosen by CDF
\cite{CDF}), and $\pi_{1,\perp}=\pi_{2,\perp}=0;9.47$ (latter cut
corresponding to a representative 1\,GeV infrared cut on both transverse momenta in the muon
pair analyzed by CDF).
Notice how severely the kinematic cuts on $\pi_{i,\perp}$ and
$\cos\theta_0$ influence the spectral shape in view of the
localization of maxima and their widths. Notice also that for
SU(2)$_\mu$ the unconstrained case ($\pi_{i,\perp}=0$ and
$\cos\theta_0=-1$) corresponds to a spectral maximum of about 0.6\,GeV
while for $\pi_{i,\perp}=9.47$ and $\cos\theta_0=-1$ the maximum
shifts to about 2.3\,GeV.

Setting $\sqrt{s}=1\,$TeV (a TeV photon generated at the Tevatron and
depositing its energy into an SU(2)$_\mu$ hot-spot by virtue of the
CDF detector material) and $m_{l=1}=m_\mu=105.6\,$MeV, one has
\eab
\label{csexp2}
c_{2(l=1);y=1,\delta=1/2}&=&7.6\times 10^4;\ \ c_{2(l=1);y=1,\delta=1}=3.04\times 10^5;\nonumber\\
c_{2(l=1);y=1/2,\delta=1/2}&=&2.29\times 10^7;\ \
c_{2(l=1);y=1/2,\delta=1}=9.14\times 10^7\,.  \eae
For example, considering SU(2)$_\mu$, $\sqrt{s}=1\,$TeV, and
$\delta=1$ Eq.\,(\ref{csexp2}) and Fig.\,\ref{Fig-2} imply that the
number $N^{\tiny\mbox{max}}_{2(l=1)}$ of muon pairs (regardless of the
charge of their participants) emitted at the spectral maximum per
invariant-mass bin of $m_\mu$ is
\eab
\label{maxnumb2}
N^{\tiny\mbox{max}}_{2(l=1)}(\delta=1,y=1,\pi_{i,\perp}=0,\cos\theta_0=-1)&=&3.04\times
10^5\times 0.44=1.34\times 10^5\,,\nonumber\\
N^{\tiny\mbox{max}}_{2(l=1)}(\delta=1,y=1,\pi_{i,\perp}=0,\cos\theta_0=0.8)&=&3.04\times
10^5\times 0.07=2.13\times 10^4\,,\nonumber\\
N^{\tiny\mbox{max}}_{2(l=1)}(\delta=1,y=1,\pi_{i,\perp}=9.47,\cos\theta_0=0.8)&=&3.04\times
10^5\times 4\times 10^{-4}=122\,.\nonumber\\ \eae
A comparison of Eq.\,(\ref{maxnumb2}) with the experimental counts of
dimuon events per bin invariant mass \cite{CDF} can be used to obtain
an estimate for the total number of hot-spots created within a given
value of total integrated luminosity in TeV-range collider
experiments.

\section{Summary and Conclusions\label{S&C}}

Concerning the data analysis in TeV-range collider experiments, the
lesson of the present article is that certain prejudices (kinematic
cuts) applied to experimental data in extrapolating known physics (the
SM) into the unknown necessarily hides potentially important
signatures. The possibility that metastable hot-spots of a new phase
of matter are not only created in ultrarelativistic heavy-ion
collisions (RHIC), where due to Quantum Chromodynamics they are
expected, but also in isolated, TeV-range proton-antiproton or
proton-proton collisions suggests a scenario completely different from
the perturbative SM approach for the deposition of center-of-mass
energy $\sqrt{s}$ into collision products. Rather than creating
secondaries of energy comparable to $\sqrt{s}$, energy would then be
re-distributed on a large multiplicity of charged, low-energy leptons
stemming from hot-spot evaporation. Because of the missing confinement
mechanism in products of SU(2) Yang-Mills theories with equal
electric-magnetic parity \cite{Hofmann2005} it is clear that only the
leptonic sector exhibits these large multiplicities explicitely. In
view of the large data stream that will be generated by the LHC we
hope that the present work will contribute to the process of
overcoming these prejudices.

An exception to this rule is the deposition of $\sqrt{s}$ into the
mass of intermediate vector bosons propagating in the preconfining
phase of SU(2) Yang-Mills theory. A genuine prediction here is that
two more but much heavier copies of vector-boson triplets must
eventually be excited.  For SU(2)$_\mu$ the mass of the neutral boson
$Z^\prime$ is hierarchically larger then $m_\mu$ (since this vector
mode decouples at $T_{\tiny\mbox{H}}\sim m_\mu$). In this context, it
is worth pointing out the potential resonance peak at $M\sim 240\,$GeV
(hierarchy $M/m_\mu\sim 2\times 10^3$) seen in the $e^+ e^-$ invariant
mass spectrum generated at Tevatron Run II and reported by both
collaborations CDF \cite{CDFZ} and D0 \cite{DOZ}. Notice that
resonances in this channel are practically free of hadronic
contaminations. If true then this resonance is a plausible candidate
of the decoupling dual gauge-field mode $Z^\prime$ of SU(2)$_\mu$
\cite{Hofmann2005}.

Concerning the interpretation of narrow-width and correlated single
electron or positron peaks in supercritical heavy-ion collisions, we
believe that the present article has established a plausible
connection to the creation of SU(2)$_e$ preconfining hot-spots and
their subsequent evaporation.

\end{document}